\documentclass[12pt,a4paper]{article}
\usepackage{latexsym}	
\usepackage{amsmath}
\usepackage{amsthm}
\usepackage{amssymb}
\usepackage{graphicx}

\usepackage{amsmath}
\usepackage{amsthm}
\usepackage{amsmath,amssymb}
\usepackage{graphicx}
\def\ba{\begin{eqnarray}}
\def\ea{\end{eqnarray}}

\def\ba{\begin{eqnarray}}
\def\ea{\end{eqnarray}}

\def\be{\begin{equation}}
\def\ee{\end{equation}}

\theoremstyle{plain}

\begin{document}
\title{LSZ ghostbusters in the quadratic gravity stage}

\author{ Osvaldo P. Santill\'an$^1$
\thanks{Electronic addresses: firenzecita@hotmail.com}\\
\textit{\small{$^1$Instituto de Matem\'atica Luis Santaló (IMAS) and CONICET,}}\\
\textit{\small{Ciudad Universitaria, 1428 Buenos Aires, Argentina.}}}
\date{}
\maketitle
\begin{abstract}
The present letter considers the quantization method developed in \cite{strumia1}-\cite{strumia3}, which postulates that, in several situations, negative norm or ghost states can be avoided in order to give positive probabilities. These authors also postulate a candidate for a path integral for those theories, following pioneer works initiated by Dirac \cite{dirac} and Pauli \cite{pauli}. It is of interest the derivation of the LSZ rules in this context, since it may not be clear at first sight that has the usual form of the textbooks, due to the non standard oscillator algebra and the redefinitions of the states. This is done here, applied to Stelle gravity \cite{stelle1}-\cite{stelle2}. In addition, an equivalent but simpler way to deal with negative norm states is worked out, in which hermiticity is more explicit. As far as we understand, our conclusions fully agree with the arguments and expectations of those authors. 
\end{abstract}

\section{Introduction}

Gravity theories  are  elusive to be quantized. There are mechanisms which are close to achieve several features that a consistent gravity model has to have. However, there is always some important points which invalidates the model. Largely studied  examples are the quadratic models of gravity, which posses fourth order equations of motion and were considered as a plausible quantum gravity candidates in \cite{dewitt}. These models were studied further by Stelle in \cite{stelle1}-\cite{stelle2}, in the context of modern QFT.  These classic references, in particular \cite{stelle2} consider the path integral quantization of theories, the problem of gauge fixing and the solution of the Slavnov-Taylor identities for these scenarios.  These theories are renormalizable, and the general form of the finite counterterms have been already worked out.

The authors of  \cite{stelle1}-\cite{dewitt} recognize that these theories apparently contain a ghost in the physical spectrum, which may indicate instabilities and unitarity loss \cite{pais}. The problem is that  propagator of such four order theory will be given in a Pauli Villars form as
$$
D\sim \frac{1}{k^2}-\frac{1}{k^2+M^2}.
$$
For  large values of $k$ this propagator will go as $M^2 k^{-4}$. In QFT
the mass $M$ is interpreted as a cutoff, which tends $M\to\infty$ at the end of the calculations. In the gravity approach, this cutoff  is not taken to be infinite, and the minus sign in the last propagator expression 
suggests the presence of an state with negative kinetic energy, which may indicate a potential instability. 

Despite this problem, several authors \cite{mannheim1}-\cite{donogue} have made considerable effort in order to interpret those states, instead of abandon the theory. These references suggest that the problem of the apparent instabilities may be solved by employing variants of the quantization method, some of them inspired in earlier works of  Dirac and Pauli \cite{dirac}-\cite{pauli}. The philosophy is to develop a quantization method which reduces to the ordinary one if no ghosts are present, and which avoid the problem of negative norm states if ghost appear. These procedures keep alive the interest in the quadratic gravity models since, if the ghost problem is worked out, they will become a promising candidate for a quantum gravity model. 

In the present letter, we take the chance of applying one of these quantization methods, those developed in \cite{strumia1}-\cite{strumia3} to derive the LSZ reduction formulas for the model. These authors figure out the expression for the Feynman path integral for calculating correlation functions. However, due to the non standard oscillator algebra one may suspect that perhaps, once the LSZ reduction is achieved, an accident may appear and non positive probabilities emerge again.  For instance, the standard formula for an scalar field $S$ matrix is
$$
S=:e^{\int \phi_{1 in}(z)\square_z\frac{\delta}{\delta J(z)}d^4z} :Z(J)\bigg|_{J=0},
$$
and even if $Z(J)$ is interpreted as a vacuum transition probability, a change in the exponential from imaginary to real or viceversa, may generate a decaying behaviour which spoils the $S$-matrix formalism.
The deduction presented along the text shows that luckily, this accident do not happen and consistent scattering cross sections may be constructed by applying  the recipes of \cite{strumia1}-\cite{strumia3}.  Even though this, one has to track the changes of the LSZ rules when these strange states are included. At the end, this changes are minimal, but this checking is mandatory.

The present work is organized as follows. The second section  is devoted to a quantization scheme and LSZ reduction  which, in authors opinion, is a simplified version of the arguments of \cite{strumia1}-\cite{strumia3}, less systematic but allowing to visualize some results in simple manner. The third section is devoted to the same problem but with the full machinery of \cite{strumia1}-\cite{strumia3}. Both results are related to the Feynman path integral, the last one is given in terms of a functional form for the $S$-matrix. The Feynman generating functional is analyzed in the fourth section. The fifth and last section contains the discussions of the presented results. 

\section{LSZ formulas for quadratic gravity}
\subsection{The modes of the model}
The action for quadratic model considered by Stelle \cite{stelle1}-\cite{stelle2} can be expressed as
$$
S=-\frac{1}{2\kappa}\int \sqrt{-g} d^4x\bigg[R+\frac{\gamma}{2} R^2-\frac{\alpha}{2}C_{\mu\nu\alpha\beta}C^{\mu\nu\alpha\beta}\bigg].
$$
Here  $C_{\mu\nu\alpha\beta}$  represents the Weyl tensor, which is constructed in terms of the scalar curvature  $R$,  the Ricci tensor $R_{\mu\nu}$ and the curvature tensor  $R_{\mu\nu\alpha\beta}$.  The standard expression of the lagrangian employed in the literature is given in terms of  $R$, $R^2$ and $R_{\mu\nu}R^{\mu\nu}$, and the difference with the above lagragian is a total derivative. The advantage of the action written in terms of the Weyl tensor is that the mass of the modes will be expressed in terms of the parameters $\gamma$ and $\alpha$ in simpler manner. 

The Stelle model is known to be renormalizable \cite{stelle1}, but it contains apparent instabilities in form of ghosts. The task is to study this problem in more detail, following \cite{strumia1}-\cite{strumia3}.

Consider a small perturbation around the Minkowski metric
$$
g_{\mu\nu}=\eta_{\mu\nu}+(\widetilde{h}_{\mu\nu}-\frac{1}{2}\eta_{\mu\nu}\widetilde{h}).
$$
This perturbation may  be decomposed as follows \cite{medeiros1}-\cite{medeiros2}
$$
\widetilde{h}_{\mu\nu}=h_{\mu\nu}+\Psi_{\mu\nu}+\frac{\eta_{\mu\nu}}{2}\bigg(\phi+\frac{\Psi}{3}\bigg).
$$
Here $h_{\alpha\beta}$ is the solution of the GR equations of motion in the traceless transverse gauge, which will be specified below. The component $\Psi_{\mu\nu}$ is of spin  two and the  field $\phi$ is an scalar.  The last expression for the perturbation assumes that the  trace of the perturbation is non necessarily zero, and its value is encoded in $\phi$. The curvature tensor for a generic perturbation can be expanded by the formulas
$$
\Gamma_{\rho\mu\nu}^{\alpha}=\frac{1}{2}\eta^{\alpha\beta}(\partial_{\mu} \widetilde{h}_{\beta\nu}+\partial_{\nu}\widetilde{h}_{\beta\mu}-\partial_{\beta}\widetilde{h}_{\mu\nu})+O(h^2),
$$
$$
R^\alpha_{\rho\mu\nu}=\frac{1}{2}\eta^{\alpha\beta}(\partial_\rho\partial_{\mu}\widetilde{h}_{\beta\nu}-\partial_\rho\partial_{\nu}\widetilde{h}_{\beta\mu}-\partial_\nu\partial_{\beta}\widetilde{h}_{\rho\mu}+\partial_\mu\partial_{\beta}\widetilde{h}_{\rho\nu})+O(h^2),
$$
and $R_{\mu\nu}=\eta^{\alpha\beta}R_{\mu\alpha\nu\beta}$, $R=\eta^{\alpha\beta}R_{\alpha\beta}$ up to higher orders in  $\widetilde{h}^{\mu\nu}$.  The determinant can be expanded as
$$
\sqrt{-g}=\sqrt{\det(-\eta- \widetilde{h})}=\exp{\frac{1}{2}\log(\det(-\eta-\widetilde{h}))}=\exp{\frac{1}{2}\log(\det(-1-\eta \widetilde{h}))},
$$
the last step employs that $\eta_{\mu\nu}=\eta^{-1}_{\mu\nu}$. By taking into account the identity "$\log \det=\text{Tr}\log$" it follows that
$$
\sqrt{-g}=\exp{\frac{1}{2}\text{Tr}(\log(-1-\eta \widetilde{h}))},
$$
and by a simple Taylor expansion of the exponential  and the logarithm it is found that
$$
\sqrt{-g}\simeq 1+\frac{1}{2} \widetilde{h}_\alpha^\alpha+\frac{1}{8} \widetilde{h}_\alpha^\alpha \widetilde{h}_\beta^\beta-\frac{1}{4} \widetilde{h}_\alpha^\beta  \widetilde{h}_\beta^\alpha.
$$
With the help of these expansions,  the lagrangian may be linealized and it becomes
$$
L=\frac{1}{2\kappa}\bigg[\frac{1}{4}\partial _{\mu}h^{\alpha \beta }\partial ^{\mu }h_{\alpha \beta }+\frac{1}{4}\partial _{\mu }\Psi^{\alpha \beta }\partial ^{\mu }\Psi_{\alpha \beta}-\frac{3}{2}\partial _{\mu} \phi\partial ^{\mu }\phi+\frac{1}{2}\partial _{\mu}h^{\alpha \beta }\partial ^{\mu }\Psi_{\alpha \beta }
$$
$$
+\partial _{\beta }h^{\alpha \beta }\partial _{\alpha }\phi
+\partial _{\beta }\Psi ^{\alpha \beta }\partial _{\alpha }\phi
. +\frac{1}{2}\partial _{\mu }h\partial ^{\mu }\phi 
\frac{1}{2}\partial _{\alpha }\phi \partial ^{\alpha }\Psi \bigg]  
$$
$$
+\frac{\gamma }{2\kappa }\bigg[-\frac{9}{2}\left( \square \phi \right)
^{2}+3\partial _{\alpha }\partial _{\beta }h^{\alpha \beta }\square
\phi +3\partial _{\alpha }\partial _{\beta }\Psi ^{\alpha \beta }\square \phi 
+\frac{3}{2}\square h\square \phi   -\frac{3}{2}\square \Psi \square \phi \bigg]  
$$
$$
+\frac{\alpha }{2\kappa }\left[ \frac{1}{4}\square h_{\alpha \beta
}\square h^{\alpha \beta }+\frac{1}{2}\square h_{\alpha
\beta }\square \Psi ^{\alpha \beta } +\frac{1}{4}\square \Psi _{\mu \beta }\square \Psi ^{\mu \beta }%
\right]+L_{int},
$$
up to interaction terms collected in $L_{int}$ that appear due to the higher order expansion of the curvature and the metric determinant. These contributions generate an infinite set of interaction vertices whose effect may be cancelled by a finite type of counterterms, as the theory is renormalizable  \cite{stelle1}, \cite{dewitt}.

The  resulting free equation of motion for $\phi$  follows from this linearization  by ignoring the term $L_{int}$. The  Euler-Lagrange equation of motion is then
\begin{equation}\label{phi}
3\square\phi- \partial_{\alpha\beta} h^{\alpha\beta}- \partial_{\alpha\beta} \Psi^{\alpha\beta}- \frac{1}{2}\square h+\frac{1}{2}\square \Psi=\frac{\gamma}{2}\bigg[18\square^2 \phi-6\square \partial_\mu \partial_\nu h^{\mu\nu}-6\square \partial_\mu \partial_\nu \Psi^{\mu\nu}-3\square^2 h+3\square^2 \Psi\bigg],
\end{equation}
while for the spin two perturbation $\Psi_{\mu\nu}$ the equations become
\begin{equation}\label{psi}
\square \Psi_{\mu\nu}+\square  h_{\mu\nu}+2\square \phi-\eta_{\mu\nu}\square \phi-6\gamma \square \partial_\mu\partial_\nu \phi+3\gamma\eta_{\mu\nu}\square^2\phi-\alpha \square^2 h_{\mu\nu}-\alpha \square^2 \Psi_{\mu\nu}=0.
\end{equation}
The transverse-traceless gauge choice mentioned above is not only applied for $h_{\mu\nu}$ but for $\Psi_{\mu\nu}$ as well. This gauge is defined by 
$$
\partial_\nu h^{\mu\nu}=\partial_\nu \Psi^{\mu\nu}=0,\qquad  h=\Psi=0,
$$
and it reduces the equation for the scalar perturbation  to
$$
\square^2 \phi-\frac{1}{3\gamma}\square\phi=0,
$$
which shows that $\phi=\phi_1+\phi_2$ is a free real scalar field with two modes $\phi_1$ and $\phi_2$ with masses $m^2_1=0$ and $m^2_2=m^2=\frac{1}{3\gamma}$. On the other hand, in the equation of motion of $\Psi_{\mu\nu}$ given by \eqref{psi}, the terms proportional to $\phi$ in \eqref{psi} cancel in pairs due the last equation. Furthermore, as $h_{\mu\nu}$ was chosen as the massless GR solution in the transverse-traceless gauge, it follows that $\square h_{\mu\nu}=0$. These simplifications reduce the equations of motion \eqref{psi} to
$$
 \square^2 \Psi_{\mu\nu}-\frac{1}{\alpha}\square \Psi_{\mu\nu}=0,
$$
which shows that $\Psi_{\mu\nu}=\Psi^1_{\mu\nu}+\Psi_{\mu\nu}^2$ is a massive spin two field with two excitations $\Psi^1_{\mu\nu}$ and $\Psi^2_{\mu\nu}$, one massless and one with mass $M^2=\frac{1}{\alpha}$.

The obtained equations impliy the following mode expansion for the fields. The graviton can be expressed as
$$
h_{\mu\nu}(x)=\int\frac{d^3k}{(2\pi)^{\frac{3}{2}}\omega(k)}[a_{\mu\nu}(k) e^{i\omega(k) t-ik x}+a^\dag_{\mu\nu}(k) e^{-i\omega(k) t+ik x}],
$$
where the coefficients are symmetric and due to the gauge condition they satisfy $$a_{\mu\nu}(k)\eta^{\mu\nu}=0,\qquad k^i a_{\mu i}(k)=\omega a_{\mu 0}(k).$$ The dispersion relation is the massless one $\omega(k)=|k|$.  For the scalar mode $\phi$ 
$$
\phi(x) =\int\frac{d^3k}{(2\pi)^{\frac{3}{2}}\omega(k)}[b(k) e^{i\omega(k)t-i k\cdot x}+b^\ast(k) e^{-i\omega(k)t+i k\cdot x}]
$$
$$
+\int\frac{d^3k}{(2\pi)^{\frac{3}{2}}\omega_2(k)}[c(k) e^{i\omega_2(k)t-i k\cdot x}+c^\ast(k) e^{-i\omega_2(k)t+i k\cdot x}],
$$
where the new energy is given by $\omega_2(k)=\sqrt{k^2+m^2}$. For the spin two perturbation the mode expansion becomes
$$
\Psi_{\mu\nu}(k) =\int\frac{d^3k}{(2\pi)^{\frac{3}{2}}\omega(k)}[d_{\mu\nu}(k) e^{i\omega(k)t-i k\cdot x}+d_{\mu\nu}^\ast(k) e^{-i\omega(k)t+i k\cdot x}]
$$
$$
+\int\frac{d^3k}{(2\pi)^{\frac{3}{2}}\omega_3(k)}[e_{\mu\nu}(k) e^{i\omega_3(k)t-i k\cdot x}+e_{\mu\nu}^\ast(k) e^{-i\omega_3(k)t+i k\cdot x}],
$$
where $\omega_3(k)=\sqrt{k^2+M^2}$ and
$$d_{\mu\nu}(k)\eta^{\mu\nu}=e_{\mu\nu}(k)\eta^{\mu\nu}=0,\qquad k^i d_{\mu i}(k)=\omega d_{\mu 0}(k).\qquad k^i e_{\mu i}(k)=\omega_3 e_{\mu 0}(k).$$
The next task is to study the apparent ghost in the Stelle model and to see if the instabilities may be avoided after quantization. 
\subsection{Canonical quantization of the model}
In order to find the canonical quantization of the Stelle model, it is needed to go to the hamiltonian formalism. Consider first the quantization of the scalar mode $\phi$. By taking into account the gauge conditions, the  part of the lagrangian that involves the scalar field is 
$$
L=-\frac{3}{2}\partial_\mu \phi \partial^\mu \phi-\frac{9\gamma}{2}(\square\phi)^2.
$$
This is up to terms that vanish when the transversal traceless conditions are taken into account. As the theory is of higher order, the procedure of doubling the number of variables can be applied, and there will be two coordinates and two momenta involved instead of one. The coordinates will be chosen as
\begin{equation}\label{coordenadas}
Q_1=\beta\phi, \qquad Q_2=\lambda\dot{\phi},
\end{equation}
with $\beta$ and $\lambda$ constant factors. In fact, the quantization to be employed now may allow complex values for these constants and still give real mean values for these operators \cite{strumia3}. This method, maybe ackward at first sight, will be described below and has roots from  Dirac and Pauli times \cite{dirac}-\cite{pauli}.

The canonical momenta corresponding to \eqref{coordenadas} can be found by the following formula, valid for a higher order formalism
$$
 P_1=\frac{\partial L}{\partial \dot{Q_1}}-\frac{d}{dt}\bigg(\frac{\partial L}{\partial \ddot{Q_1}}\bigg)=\frac{1}{\beta}[3\dot{\phi}-9\gamma\square \dot{\phi}],
$$
\begin{equation}\label{momentum}
 P_2=\frac{\partial L}{\partial \dot{Q_2}}-\frac{d}{dt}\bigg(\frac{\partial L}{\partial \ddot{Q_2}}\bigg)=\frac{9\gamma}{\lambda} \square \phi,
\end{equation}
while the hamiltonian density is found by the standard prescription
$$
H=P_1 \dot{Q}_1+P_2\dot{Q}_2-L.
$$
The above procedure ensures that the hamiltonian equations of motion are equivalent to the Euler-Lagrange ones. It is useful  to deduce from the second \eqref{momentum} that
$$
\ddot{\phi}=\Delta \phi-\frac{\lambda}{9\gamma} P_2.
$$
The hamiltonian density can be calculated explicitly in terms of the canonical coordinates \eqref{coordenadas} and momentum \eqref{momentum}, the result is
\begin{equation}\label{canon}
H=\frac{\beta P_1 Q_2}{\lambda}-\frac{\lambda^2 P_2^2}{18\gamma}+\frac{\lambda P_2\Delta Q_1}{\beta}-\frac{3}{2}\frac{Q_2^2}{\lambda^2}+\frac{3}{2\beta^2}(\nabla Q_1)^2.
\end{equation}
If the coefficients $\beta$ and $\lambda$ are chosen equal to one, then the euclidean path version of this theory  wlll be divergent since the integration on the variable $P_1$ yields an infinite result. This follows from the fact that the hamiltonian is linear in $P_1$. This is not the case for other choices of these constants, for instance imaginary ones. These choices do not affect the canonical commutation relations $[P, Q]$ as they multiply $Q$ in inverse way as for $P$. The point is to understand the consequences of declaring that $\lambda$ or $\beta$ are imaginary. Related work is the reference \cite{kleefeld}, where  apparently a self-consistent way to quantize these systems has been presented.

The quantization method requires to  express the mode coefficients $b(k)$, $b^\ast(k)$, $c(k)$ and $c^\ast(k)$  in terms of  the canonical variables $P^i$ and $Q_i$ with $i=1,2$. Then by assuming the standard equal time commutation relations given by
\begin{equation}\label{canon}
[Q_i(x,t), P_j(y, t)]=i\hbar \delta_{ij} \delta(x-y).
\end{equation}
the commutation relations for the quantum operators $b(k)$, $b^\dag(k)$, $c(k)$ and $c^\dag(k)$ can be deduced. This is a $4\times 4$ system of equations.   As it will be shown below, it is not possible to enforce this canonical equal time commutators \eqref{canon}  if the coordinates commute $$[Q_i(x,t), Q_j(y,t)]=0.$$ This fact leads to the negative norm states, which will be avoided with imaginary choices of $\beta$ and $\lambda$.

A convenient simplification occurs by decomposing $$\phi=\phi_1+\phi_2,$$ with $\phi_1$ the massless mode and $\phi_2$ the one with mass $m^2=\frac{1}{3\gamma}$. The second of the formulas \eqref{momentum} shows that this massless mode does not contribute to $P_2$ as $\square \phi_1=0$. Furthermore, from  the first \eqref{momentum} and the definition $\dot{\phi}=\lambda Q_2$
it is seen that it does not contribute to $\beta P_1-\frac{3 Q_2}{\lambda}$ either.  Therefore, only the massive mode $\phi_2$ contributes to these two quantities and since the massive mode involves only the coefficient $c(k)$ and $c^\ast(k)$, the system
\begin{equation}\label{ecua}
\lambda \beta P_1-3Q_2=-9\lambda\gamma\square \dot{\phi}_2,\qquad P_2=\frac{9\gamma}{\lambda}\square \phi_2,
\end{equation}
is a $2\times 2$ one allowing the calculation  $c(k)$ and $c^\ast(k)$ in terms of the modes of $\lambda \beta P_1-3 Q_2$ and $P_2$. This makes the problem more tractable. To solve it explicitly write 
$$
\phi_2(x,t)=\int\frac{d^3k}{(2\pi)^{\frac{3}{2}}\omega_2(k)}[c(k) e^{i\omega_2(k)t-i k\cdot x}+c^\ast(k) e^{-i\omega_2(k)t+i k\cdot x}].
$$
By taking into account that $\square \phi_2=m^2 \phi_2$ with $m^2=\frac{1}{3\gamma}$ the equations \eqref{ecua} are converted into
$$
\lambda P_2(x,t)=3 \phi_2(x,t)=3\int\frac{d^3k}{(2\pi)^{\frac{3}{2}}\omega_2(k)}[c(k) e^{i\omega_2(k)t-i k\cdot x}+c^\ast(k) e^{-i\omega_2(k)t+i k\cdot x}],
$$
\begin{equation}\label{ecuador}
3Q_2(x,t)-\lambda \beta P_1(x,t)=3\dot{\phi}_2=3i\int\frac{d^3k}{(2\pi)^{\frac{3}{2}}}[c(k) e^{i\omega_2(k)t-i k\cdot x}-c^\ast(k) e^{-i\omega_2(k)t+i k\cdot x}].
\end{equation}
This can be inverted to give
$$
c(k)=\frac{1}{3}\int d^3x e^{ikx}[\omega_2(k)\lambda P_2(x,0)-3iQ_2(x,0)+i\lambda\beta P_1(x,0) ],
$$
\begin{equation}\label{oscila}
c^\dag(k)=\frac{1}{3}\int d^3x e^{-ikx}[\omega_2(k)\lambda P_2(x,0)+3iQ_2(x,0)-i\lambda\beta P_1(x,0) ].
\end{equation}
When these expressions are prompted to operators $\widetilde{c}_k$ and $\widetilde{c}_k^\dag$, by assuming the canonical commutation relations \eqref{canon} it is arrived to
\begin{equation}\label{lulu1}
[\widetilde{c}_{k^\prime}, \widetilde{c}^\dag_k]=\frac{2\lambda\omega_2(k)}{3}\delta(k-k^\prime),
\end{equation}
This is the standard commutation relation up to a factor $3\lambda^{-1}$.
Consider now the remaining operators $b(k)$ and $b^\dag(k)$. Their commutation relation  can be determined from \eqref{coordenadas}, as $[Q_1, Q_2]=0$.  This definition implies that
$$
Q_1 =\beta\int\frac{d^3k}{(2\pi)^{\frac{3}{2}}\omega(k)}[b(k) e^{i\omega(k)t-i k\cdot x}+b^\ast(k) e^{-i\omega(k)t+i k\cdot x}]
$$
$$
+\beta\int\frac{d^3k}{(2\pi)^{\frac{3}{2}}\omega_2(k)}[c(k) e^{i\omega_2(k)t-i k\cdot x}+c^\ast(k) e^{-i\omega_2(k)t+i k\cdot x}],
$$
$$
Q_2 =i\lambda \int\frac{d^3k}{(2\pi)^{\frac{3}{2}}}[b(k) e^{i\omega(k)t-i k\cdot x}-b^\ast(k) e^{-i\omega(k)t+i k\cdot x}]
$$
$$
+i\lambda\int\frac{d^3k}{(2\pi)^{\frac{3}{2}}}[c(k) e^{i\omega_2(k)t-i k\cdot x}-c^\ast(k) e^{-i\omega_2(k)t+i k\cdot x}].
$$
As these two quantities commute, when the unknowns $b_k$ and $b^\ast(k)$ are prompted to operators $\widetilde{b}_k$ and $\widetilde{b}_k^\dag$ the commutation relation $[\widetilde{b}_k,\widetilde{b}_k^\dag]$ has to cancel the terms coming from $[c_k,c_k^\dag]$. This leads to
\begin{equation}\label{commutacion2}
[b_k,b_{k^\prime}^\dag]=-\frac{2\lambda\omega_2(k)}{3}\delta(k-k^\prime).
\end{equation}
In other words, one of the two pair of creation/annihilation operators has a non standard $-1$ sign in the commutation relation in order to ensure that $[Q_1, Q_2]=0$.  This  non standard sign is unavoidable. The last commutation relation leads to an expression of $[Q_1, P_1]$ at equal times to have a minus sign in their definition, that is
$$
[Q_1(x,t), P_1(y,t)]=-i\hbar\delta(x-y).$$ 
One may declare a non standard minus sign in $[Q_2, P_2]$ instead, and the commutation $[Q_1, P_1]$ will be standard.   But it is not possible to impose \eqref{canon} for both pairs of canonical variables, at least if $[Q_1, Q_2]=0$. It is customary in the literature to impose the non standard commutation relation for the massive mode, since if its mass is extremely large, its effect is softened. As the text below is intended to deal with negative norm states, this convention will not play a role.

The hamiltonian  in terms of $\phi$ is given by
$$
H=[3\dot{\phi}-9\gamma\square \dot{\phi}]\dot{\phi}+9\gamma \square \phi\ddot{\phi}+\frac{3}{2}\partial_\mu \phi \partial^\mu \phi+\frac{9\gamma}{2}(\square\phi)^2.
$$
By decomposing $\phi=\phi_1+\phi_2$ and by taking into account their equations of motion, it is seen that the terms mixing $\phi_1$ and $\phi_2$ cancel out, and the resulting hamiltonian density is simply the sum of two free Klein-Gordon ones corresponding to both modes. The calculation leading to this conclusion goes as follows. The mixed terms are 
$$
H_{12}=[3\dot{\phi}_1-9\gamma\square \dot{\phi}_1]\dot{\phi}_2+9\gamma \square \phi_1\ddot{\phi}_2+3\partial_\mu \phi_1 \partial^\mu \phi_2+9\gamma\square\phi_1\square \phi_2
$$
$$
+[3\dot{\phi}_2-9\gamma\square \dot{\phi}_2]\dot{\phi}_1+9\gamma \square \phi_2\ddot{\phi}_1.
$$
From the equations of motion $\square\phi_1=0$ and $3\gamma\square\phi_2=\phi_2$ it is clear that a lot of simplifications occur, and the last expression becomes 
$$
H_{12}=3\dot{\phi}_1\dot{\phi}_2+3\partial_\mu \phi_1 \partial^\mu \phi_2
+3\phi_2\ddot{\phi}_1.
$$
Collecting the first and the last term as a total derivative, it follows that
$$
H_{12}=3\partial_t(\dot{\phi}_1\phi_2)-3\dot{\phi}_1\dot{\phi}_2+3\nabla \phi_1 \cdot\nabla \phi_2.
$$
This can be equivalently written as
$$
H_{12}=3\phi_2\ddot{\phi}_1+3\nabla \phi_1 \cdot\nabla \phi_2.
$$
This is a mixture. However, after integrating this density over the whole spatial volume, the last term is equivalent by parts integration to 
$$
H_{12}=3\phi_2\ddot{\phi}_1-3\phi_2\Delta \phi_1.
$$
Since $\ddot{\phi}_1\sim -\omega^2 \phi_1$ while $\Delta \phi_1\sim -k^2\phi_1$ inside the mode expansion, the dispersion relation for $\phi_1$ $\omega^2=k^2$  implies that the contribution of these terms can be neglected.  The resulting hamiltonian density is
$$
H=\frac{3}{2}\dot{\phi}^2_1+\frac{3}{2}(\nabla\phi_1)^2+3\phi_2\ddot{\phi}_2-\frac{3}{2}\dot{\phi}^2_1+\frac{3}{2}(\nabla\phi_2)^2+\frac{3m^2}{2}(\phi_2)^2.
$$
Inside the Fourier expansion $\ddot{\phi}_2=\dot{\phi}_2^2\sim -\omega_2^2 \phi_2$, thus the last expression is equivalent to
$$
H=\frac{3}{2}\dot{\phi}^2_1+\frac{3}{2}(\nabla\phi_1)^2+\frac{3}{2}\dot{\phi}^2_1+\frac{3}{2}(\nabla\phi_2)^2+\frac{3m^2}{2}(\phi_2)^2.
$$
This clearly the sum of two free  Klein-Gordon hamiltonian densities, for a massless and a massive particle.  By omitting the zero point energy, the hamiltonian written in terms of the modes is given by
$$
H=\int d^3k[ \omega(k)b^\dag(k)b(k)+ \omega_2(k)c^\dag(k)c(k)].
$$
One of the problems to be discussed now is how to avoid this hamiltonian to be unbounded from below.
\subsection{Interpretation of the non standard creation annihilation algebra}

In order to analyze the consequences of the  non standard  creation annihilations algebra obtained above, consider a simple quantum mechanical oscillator with creation $u_c$ and annihilation $u_a$ operators satisfying
\begin{equation}\label{lolita}[u_a,u_c]=-1.\end{equation} 
These operators may be interpreted heuristically as given by the standard creation an annihilation operators multiplied by $i$, that is $u_a=ia$, $u_c=ia^\dag$.   Define the standard  $n$ quanta state
$$
|n>=\frac{a^\dag}{\sqrt{n}}|n-1>=\frac{(a^\dag)^n}{\sqrt{n!}}|0>,
$$
such that
$$
a|n>=\sqrt{n}|n-1>.\qquad a^\dag|n>=\sqrt{n+1}|n+1>.
$$
In other words, these states are the ones constructed with the standard version of the creation annihilation algebra. The last formula implies that
\begin{equation}\label{ifactor}
u|n>=i\sqrt{n}|n-1>.\qquad u^\dag|n>=i\sqrt{n+1}|n+1>.
\end{equation}
Here of course $$[a,a^\dag]=1.$$ The mean values with respect to the usual basis $|n>$ of the hamiltonian
$$
H=\omega u^\dag u,
$$
are negative if $\omega$ is positive, since
$$
<n|H|n>=-\omega n,
$$
due to the $i$ factor in \eqref{ifactor}. Furthermore, the standard creation operators can be expressed as
$$
 a=\frac{p+ix}{\sqrt{2}},
$$
and $p$ and $x$ are the usual quantum mechanical position and momentum operators. These satisfy  $[x, p]=i$ and
$$
p|x>=-i\frac{d}{dx}|x>, \qquad x|x>=x|x>.
$$
Based on this, and that $u=ia$, it may be postulated that
$$
u=\frac{\beta P+\frac{iX}{\beta}}{\sqrt{2}},
$$
leading to
$$
\beta P|x>=\frac{d}{dx}|x>, \qquad \frac{X}{\beta}|x>=ix|x>.
$$
Note that an arbitrary constant $\beta$ has been included, in the same fashion as in the definition \eqref{coordenadas}. Following the discussion below that formula chose $\beta=i$. Then 
$$
 P|x>=-i\frac{d}{dx}|x>, \qquad X|x>=-x|x>.
$$
In other words, $P$ is equal to the standard $p$ but $X$ has a wrong minus sign. Therefore $<X>=-x$
in a coordinate basis. The mean values of $P$ are standard.  If one declares that the mean values should be taken with respect $|X>=|x>$ and to $<X|=-<x|$ then the values of $X$ are corrected and the ones of $P$ change a sign, which is harmless. This second choice seems more attractive.

Note that both $P$ and $X$ are hermitian with a minus sign in the standard commutator. The advantage of the  procedure of taking mean values with the bra with a switched sign is that the mean values  of the hamiltonian become
$$
E_n=<N|H|N>=\omega n,
$$
where $<N|=-<n|$ and $|N>=|n>$. The energies are now positive defined. Therefore it is convenient to stay tuned with this procedure for calculating the mean values.

This suggest that the mean values  $<\Xi_n|O|\Xi_m>$ should be calculated by a sandwich between eigenstates  $|\xi_m>$  of some observable $\Xi$ constructed by the correspondence principle in terms of the usual
operators $p$ and $x$ and by making the identification $|\Xi_n>=|\xi_n>$ and $<\Xi_n|=-<\xi_n|$. The norm of the states is identified as
$$|\xi_n|^2=<\xi_n|\xi_n>=-<\Xi_n|I|\Xi_n>.$$

This means that the expression of the norm depends on the chosen basis.  For the standard one, the mean value is as usual, while for the one with switched values, it is given as the mean value of the operator $I=-1$ instead of $I=1$. Non trivial  unit operators are usual when apparent ghost are present. The transition probabilities from an state $|\alpha>$ to an state $|\beta>$ may be calculated with or without the minus sign, as these probabilities involve an square that do not distinguish this eventual minus sign. But in order to be consistent, it is convenient to employ the minus sign convention.

\subsection{Reduction scattering formulas for Stelle gravity}

Given the previous discussion, the $S$ matrix for a given process concerning the scalar $\phi$
can be found as follows. In the present context, the quantity of interest is the amplitude for going from an initial state to a final one 
$$
A_{f i}=<f, \text{out}|i,\text{in}>.
$$
Here $|i,\text{in}>$ or $<f, \text{out}|$ of interest are eigenstates of the momentum of the field. 

A subtle detail is the following.  In the procedure outlined just above, the mean values of the momentum are
$$
<p|P|p>=-p.
$$
Therefore the states $|p>$ for particles with unusual creation annihilation algebra are describing an state of momentum $-p$. This is not a serious problem. The only thing one should keep in mind is that, when the amplitude has been calculated, it is really describing the amplitude with the momentum of the ghost like particles changed by a sign. In addition, there may be a minus sign in the definition of $A_{fi}$ due to the conventions we are making. But this sign do not affect at all the transition probabilities.

The operator 
$$
S=\sum_{\text{all states}}I|f,\text{out}><f, \text{in}|=-\sum_{\text{all states}}|f,\text{out}><f, \text{in}|,
$$
transforms out states into in states. This means that the probability amplitude is
$$
A_{fi}=-<f, \text{in}|S|i,\text{in}>=-S_{fi}.
$$
The amplitude $A_{fi}$ is then represented by the matrix elements of $S$ between the initial and final configuration, both though as in states.  There is a minus sign that is not standard in this definition. However, this is harmless.  The transition probability from the initial to the final stated
$$
P_{fi}=S_{if} S_{i f}^{\ast}.
$$
do not distinguish this minus sign. The sum over all the final states must be 
$$
\sum_{f} P_{fi}=1.
$$
Therefore 
$$
\sum_f S_{if}S^\dag_{fi}=1.
$$
In other words the $S$ matrix is unitary
$$
 S^\dag S=1.
$$
Consider now a generic transition amplitude
$$
<p_1,.., p_l,\text{out}|r_1, .., r_n,  \text{in}>,
$$
involving only the two excitations $\phi_1$ or $\phi_2$  of the field $\phi$. Here all the variables $p_i$ or $r_i$  represent possible four momentum for these states.  From their expansion in modes
$$
\phi_1(x,t)=\int\frac{d^3k}{(2\pi)^{\frac{3}{2}}\omega(k)}[b(k) e^{i\omega_2(k)t-i k\cdot x}+b^\dag(k) e^{-i\omega_2(k)t+i k\cdot x}],
$$
$$
\phi_2(x,t)=\int\frac{d^3k}{(2\pi)^{\frac{3}{2}}\omega_2(k)}[c(k) e^{i\omega_2(k)t-i k\cdot x}+c^\dag(k) e^{-i\omega_2(k)t+i k\cdot x}].
$$
and taking into account the expression of their derivatives $\dot{\phi}_1$ and $\dot{\phi}_2$
it is an elementary Fourier  exercise to determine $b(k)$, $b^\dag(k)$, $c(k)$ and $c^\dag(k)$ in terms of the values of these quantities at $t=0$. The result is
$$
b(k)=\int d^3x e^{ikx}[\omega(k)\phi(x,0)-i\dot{\phi}(x,0)] \qquad b^\dag(k)=\int d^3x e^{-ikx}[\omega(k)\phi(x,0)+i\dot{\phi}(x,0)],
$$
and analogous formulas are valid for $c(k)$ and $c^\dag(k)$. These formulas may be expressed in terms of the operation $u\overleftrightarrow{\partial}v=u(\partial v)-(\partial u) v$ as follows
$$
b(k)=-i\int d^3x [e^{-i(\omega(k)t-kx)}\overleftrightarrow{\partial}_0\phi ]\bigg|_{t=0},\qquad b^\dag(k)=i\int d^3x [e^{i(\omega(k)t-kx)}\overleftrightarrow{\partial}_0\phi ]\bigg|_{t=0}.
$$
A small detail is in order. The algebra of these operators has for $\lambda=1$ a non standard $3$ factor, as shown in formula \eqref{commutacion2}. Therefore it is convenient to redefine $\phi\to \sqrt{3}\phi$. In these terms the desired amplitude may be written as
$$
<p_1,.., p_l,\text{out}|r_1, .., r_n,  \text{in}>=i<p_1,.., p_l,\text{out}|b^\dag_{in}(r_1)|r_2, .., r_n,  \text{in}>
$$
\begin{equation}\label{aplique}
=-\lim_{t\to -\infty}\int d^3x [e^{i(\omega(r_1)t-r_1x)}\overleftrightarrow{\partial}_0<p_1,.., p_l,\text{out}|\phi_{1 in}(x,t)|r_2, .., r_n,  \text{in}> ]\bigg|_{t},
\end{equation}
where in the first step the fact that $b^\dag(k)|n_k>=i\sqrt{n_k+1}|n_k+1>$.
The standard trick of QFT of converting this expression in a four dimensional integral only requires the use of the mode equation $\square \phi_1=0$. The non standard commutation relation for $b(k)$ and $b^\dag(k)$ does play any role. This procedure is described for instance in pages 206-209 of  \cite{itzykson}.  First, the following formula has to be applied 
$$
\bigg(\lim_{t\to\infty} -\lim_{t\to-\infty}\bigg)\int f(x,t)d^3x=\lim_{t_i\to-\infty}\lim_{t_f\to\infty}\int_{t_i}^{t_f}dt\frac{\partial}{\partial t}\int f(x,t)d^3x.
$$
together with the identity
$$
\int d^4x \partial_0[e^{i(\omega(r_1)t-r_1x)}\overleftrightarrow{\partial}_0<p_1,.., p_l,\text{out}|\phi_{1 in}(x,t)|r_2, .., r_n,  \text{in}>] 
$$
$$
=-\int d^4x (\Delta e^{i(\omega(r_1)t-r_1x)})<p_1,.., p_l,\text{out}|\phi_{1 in}(x,t)|r_2, .., r_n,  \text{in}> 
$$
$$
+\int d^4x e^{i(\omega(r_1)t-r_1x)}\partial^2_0<p_1,.., p_l,\text{out}|\phi_{1 in}(x,t)|r_2, .., r_n,  \text{in}>] 
$$
$$
=\int d^4x  e^{i(\omega(r_1)t-r_1x)}\square_x<p_1,.., p_l,\text{out}|\phi_{1 in}(x,t)|r_2, .., r_n,  \text{in}>,
$$
the last equality follows after part integration. These two identities, if applied to \eqref{aplique} lead to the following reduction formula
$$
<p_1,.., p_l,\text{out}|r_1, .., r_n,  \text{in}>=-i\sum_{k=1}^l 2\omega(r_1)(2\pi)^{3} \delta^3(p_k-r_1)<p_1,.\hat{p}_k., p_l,\text{out}|r_1, .., r_n,  \text{in}>
$$
$$
+Z^{-\frac{1}{2}}\int d^4x e^{i(\omega(r_1)t-r_1x)}\square <p_1,.., p_l,\text{out}|\phi_1(x,t)|r_2, .., r_n,  \text{in}>,
$$
with the  hat symbol in $\hat{p}_k$ indicating that this particular momentum is missing. This reduction formula is almost the same than in QFT, up to irrelevant factors with unit modulus. The procedure  may be generalized to extract all the fields  in order to find
$$
<p_1,.., p_l,\text{out}|r_1, .., r_n,  \text{in}>=(-i)^{n+l}\text{disconnected terms}
$$
$$
+Z^{-\frac{n+l}{2}}\int d^4x_1...d^4y_n e^{i\sum_{k=1}^l(\omega(r_k)t_k-r_kx_k)} e^{-i\sum_{m=1}^n(\omega(p_m)t_m-p_mx_m)}
$$
$$
\square_{y_1}..\square_{x_n}<0|T\phi_1(y_1)...\phi_1(x_n)|0>.
$$
In this way, the LSZ reduction formulas for the apparent ghost mode $\phi_1$ is reduced to the standard one involving the correlations $<0|T\phi_2(y_1)...\phi_2(x_n)|0>$. For the other mode $\phi_2$ or mixtures of both the formula is exactly the same. A generic amplitude for the Stelle gravity reads
$$
<p_1,.., p_l,\text{out}|r_1, .., r_n,  \text{in}>=(-i)^{n+l}\text{disconnected terms}
$$
\begin{equation}\label{lsz}
+Z^{-\frac{n+l}{2}}\int d^4x_1...d^4y_n e^{i\sum_{k=1}^l(\omega(r_k)t_k-r_kx_k)} e^{-i\sum_{m=1}^n(\omega(p_m)t_m-p_mx_m)}
\end{equation}
$$
\square_{y_1}..(\square_{y_a}+m^2)..(\square_{y_b}+M^2)...\square_{x_c}..\square_{x_n}<0|T\phi_1(y_1)..\phi_2(y_a)...\Psi_{\alpha\beta}(y_b)..h_{\mu\nu}(x_c)..\phi_2(x_n)|0>.
$$
As expected, everything is found in terms of correlation functions.  

The expression obtained in \eqref{lsz} is equivalent to the standard textbook formula \cite{itzykson}, except for an imaginary overall factor which does not affect the transition probability. This affirmation has to be taken not literally yet, as the correlation functions have not been fully characterized. Before studying these correlations, it is convenient to present an alternative way of deducing LSZ formulas.

\section{LSZ formulas for quadratic gravity in the covariant-contravariant Hilbert space formalism}
\subsection{Covariant and contravariant distinction in the Hilbert space}
It is convenient to compare the presented results with literature. The authors \cite{strumia1}-\cite{strumia3} employ a formalism in which $u^\dag=a^\dag$ and $u=-a$ instead of the choice $u^\dag=ia^\dag$ and $u=ia$ made along the text. Then they calculate the eigenstates of  $u$ and $u^\dag$, instead  the standard ones  considered here. This choice allows them to show that the eigenstates wave functions are normalizable \cite{strumia3}. But the main point is that the norm of these states in \cite{strumia1}-\cite{strumia3}  are, due to their convention, $$<n|m>=\eta_{nm}=(-1)^n \delta_{nm},$$ while in the formalism presented above all the norms  were negative.  

The metric $<n|m>=\eta_{nm}=(-1)^n \delta_{nm}$,  is not defined positive. However, this happens also in Special Relativity, where the Minkowski metric is not defined positive with covariant or contravariant indices, but for mixed indices one has $\eta_\mu^\nu=\delta_\mu^\nu$, which does not have negative entries. Without trying to make a Minkowski Hilbert space, consider the possibility of having covariant and contra variant quantum states.

As for an harmonic oscillator the states with odd $n$ have odd parity, the non positive internal product corresponding to the choice in \cite{strumia3}   may be written generically as
\begin{equation}\label{say}
<\psi'|\psi> =  \int dx\, [ \psi^{\prime*}_{\rm even}(x) \psi_{\rm even}(x) -\psi^{\prime*}_{\rm odd}(x)\psi_{\rm odd}(x)]
= \int dx \, \psi^{\prime*}(x)  \psi(-x)  .  
\end{equation}
The unit operator related to this product is then
$$
1=\int dx |x><-x|.
$$
and therefore $<x|y>=\delta(x+y)$. This is different from QM where  $<x|y>=\delta(x-y)$. This is accordance with the fact obtained in the previous sections that naively $x$ has a wrong sign as eigenvalue, that is $x\to -x$. For those authors, the same happens.

The non positivity of the inner product seems problematic. However, one may define covariant and contravariant states
$$
|_n>=|n>,\qquad |^n>=\eta^{nm}|_m>,
$$
where $\eta^{nm}=\eta_{nm}$ since  $\eta^2=1$. Then it follows that while
$$
<_n|_m>=\eta_{nm},\qquad <^n|^m>=\eta^{nm},
$$
are not positive definite, the mixed inner product
$$
\qquad <_n|^m>= <^n|_m>=\delta^{m}_n,
$$
is. The last product is suited for describing probabilities. 
This positive defined product, written without expanding in a basis, is given by
$$
<\psi´|\eta|\psi>= \int dx\, [ \psi^{\prime*}_{\rm even}(x) \psi_{\rm even}(x) +\psi^{\prime*}_{\rm odd}(x)\psi_{\rm odd}(x)]
= \int dx \, \psi^{\prime*}(x)  \psi(x).
$$
This looks like the standard product taught in standard courses in QM. 

In the formalism in which there are covariant and contravariant states, it is important to distinguish between contravariant and covariant versions of the ket $|x>$. As parity under $x\to -x$ plays an important role, these states can be defined by
$$
|_x>=|x>,\qquad |^x>=|-x>,
$$
With this definition, the completeness relation can be expressed in two equivalent forms
$$
1=\int dx |_x><^x|,\qquad \eta=\int  dx |_x><_x|=\int  dx |^x><^x|.
$$
Note that depending the basis, the unit operator may acquire negative matrix elements. In the formalism of the previous section, it was completely negative. In general, this operator will not be standard in presence of ghost  like states.

Given a Hilbert space with a covariant or contravariant distinction,  an arbitrary  state $|\psi>$ can be expanded in the covariant or contravariant basis
$$
|\psi>=\psi^n|_n>=\psi_n|^n>.
$$
where $\psi^n=<^n|\psi>$ and  $\psi_n=<_n|\psi>$. In other words, a vector state $|\psi>$  has two possible representations $\psi_n$ or $\psi^n$. This  means that  the mean values of an observable $O$ have four different presentations $$O_{nm}=<_n|O|_m>,\qquad O_n^{\;m}=<n|O|^m>,\qquad O^n_{\;m}=<^n|O|_m>,\qquad O^{nm}=<^n|O|^m>.$$ In particular, this applies  for the operator $1$, which is related directly to the metric $\eta_{nm}$ or to its upper components by the following formula
$$
1=\eta_{nm}|^n><^m|= \eta^{nm}|_n><_m|=|^n><_n|.
$$
This is perfectly consistent with the completeness relations for $|_x>$ derived above.

In this formalism, it should not be expected that an hermitian operator is represented as an hermitian matrix in all the four possible presentations described above. In fact, given an operator $O$ whose action on a state is given by $|\psi^\prime>=O|\psi>$ then its adjoint operator $O^\dag$ is defined by the standard relation  $$<\psi^\prime|=<\psi| O^\dag.$$
In particular this means that 
\begin{equation}\label{conjug}
<\psi_1|O|\psi_2>=<\psi_2|O^\dag|\psi_1>^\ast.
\end{equation}
The adjoint of a matrix in all the presentations are defined by  $$O_{nm}^\dag=O_{mn}^\ast,\qquad O^{\dag mn}=O^{nm\ast}, \qquad O_{n}^{\dag m}=0^{m\ast}_{\;n}.
$$ A self adjoint operator  is defined by $O=O^\dag$.  A self adjoint operator looks hermitian for the presentations $O_{nm}$ or $O^{nm}$ but for mixed indices the components are related by the isospectral transformation
$$
O_{n}^m=(\eta O^{\ast T} \eta^{-1})_n^m.
$$
It is a remarkable fact that the  last matrix can be anti hermitian and still have real eigenvalues. It is convenient to employ the symbol $\dag$
 to denote the adjoint of an operator, but not for a matrix, in order to do not generate confusion.

The above paragraph suggest that, in this formalism, hermitian operators may have an unusual look.  An striking feature is that they may appear  to have imaginary eigenvalues while this is not being so. This remarkable fact was noticed by Dirac \cite{dirac} and Pauli \cite{pauli}. For instance, consider the coordinate $x$ and momentum $p$ operators in this formalism.  Instead of employing the usual operators for $x$ and $p$ in QM, for covariant states the operators
$$ 
\hat{x} |_x> = ix |_x>,\qquad p |_x>  = \frac{d}{dx} |_x>,
$$
should be employed for the following reasons. First, although these operators seem to have imaginary mean values due to the non standard $i$ factor in their definition, with a non positive defined inner product, is found that
$$
<_{x^\prime}|\hat{x}^\dag|_x>=<_{x^\prime}|\hat{x}|_x>^\ast=[ix^\prime\delta(x+x^\prime)]^\ast=ix\delta(x+x^\prime)=<_{x^\prime}|\hat{x}|_x>.
$$
Therefore the operator $\hat{x}$ is still self adjoint and its mean values
\begin{equation}\label{min}
<\psi|\hat{x}|\psi>=\int \psi^\ast(-x) \hat{x}\psi(x) dx.
\end{equation}
are real.  In addition, it is direct to see that $[x, p]=i\hbar$. This is the formalism employed in \cite{strumia3}.  In this formalism, the coordinates and momentum of a ghost like (or Dirac-Pauli) state are imaginary $x\to ix$, $p\to -ip$.  Our choice in the previous sections instead, was to enforce both operators to look hermitian from the beginning, but the momentum  flipped by a minus sign.

\section{LSZ reduction again, in terms of Pauli-Dirac variables}

Given the previous discussion,  consider again the $S$ matrix for a given process involving the scalar $\phi$, but now described in terms of a Hilbert space with  a covariant and contravariant state distinction. The desired amplitude is
$$
A^f_{\;i}=<^{f, \text{out}}|_{i,\text{in}}>,\qquad
S=\sum_{\text{all states}}|_{h,\text{in}}><^{h, \text{out}}|=\sum_{\text{all states}}|^{h,\text{in}}><_{h, \text{out}}|,
$$
where the mixed covariant and contra variant basis was employed, since it insures positive norm. The last two expressions mean that the probability amplitude is given in terms of the $S$ matrix by the relation
$$
A_{\; i}^f=<^{f, \text{in}}|S|_{i,\text{in}}>=S^f_{\;i},
$$
which express the transition entirely in terms of the incoming states. The operator 
$$
S^\dag=\sum_{\text{all states}}|^{h,\text{out}}><_{h, \text{in}}|=\sum_{\text{all states}}|_{h,\text{out}}><^{h, \text{in}}|
$$
is the adjoint of the operator $S$. To check this statement, note that
$$
<^{f, \text{out}}|_{i,\text{in}}>^\ast=<^{f, \text{in}}|S|_{i,\text{in}}>^\ast=<_{i, in}|^{f, \text{out}}>=<^{i, in}|_{f, \text{out}}>=<^{i, \text{in}}|S^\dag|_{f, in}>=<_{i, \text{in}}|S^\dag|^{f, in}>,
$$
from where it is read that 
$$
<^{f, \text{in}}|S|_{i,\text{in}}>^\ast=<_{i, \text{in}}|S^\dag|^{f, in}>.
$$
which is the condition defining the adjoint, thus $S^\dag$ is the desired adjoint operator. Furthermore it is clear that for normalized outgoing states
$$
1=<^{f\text{out}}|_{f\text{out}}>=\sum_{\text{all states}}<^{f\text{out}}|_{h,\text{in}}><^{h, \text{in}}|_{f\text{out}}>=\sum_{\text{all states}}<^{f\text{in}}|S|_{h,\text{in}}><^{h, \text{in}}|S^\dag|_{f\text{in}}>
$$
$$
=<^{f\text{in}}|S S^\dag|_{f\text{in}}>=<^{f\text{in}}|_{f\text{in}}>,
$$
from where it is seen that 
$$
SS^\dag=1.
$$
It can be shown that $S^\dag S=1$, which shows that the $S$ matrix when working in the mixed basis. In fact, the last relation is true for every basis, the problem is that the mean values of the operator  $1$ is a covariant or contravariant context are not ensured to be positive.

A further property of the $S$-matrix is the following. Consider the action on  the creation operator  for a state of momentum $k$ $b_{in}^\dag(k)$ defined by
$$
S^\dag a^\dag_{in}(k)S=(\sum_{\text{all states}}|_{r,\text{out}}><^{r, \text{in}}|) (\sum_{\text{all states}}\sqrt{n(k)+1}|_{h+k,\text{in}}><^{h, \text{out}}|)
$$
$$
=\sum_{\text{all states}}\sqrt{n(k)+1}|_{h+k,\text{out}}><^{h, \text{out}}|
$$
Here $n(k)$ are the number of quanta of momentum $k$ that were already present in the state $h$, and $h+k$ denotes the state with one of such quanta added. Clearly, this expression represents $b^\dag_{out}(k)$, which leads to the conclusion that
$$
S^\dag b^\dag_{in}(k)S= b^\dag_{out}(k),\qquad S^\dag b_{in}(k)S= b_{out}(k),
$$
the second relation is completely analogous to the first one. Therefore, it is deduced that
\begin{equation}\label{boca}
S^\dag \phi_{in}(k)S= \phi_{out}(k).
\end{equation}
On the other hand, the equation of motion of the ghost like part of $\phi(x)$ is given by
\begin{equation}\label{moushon}
\square \phi_1=\frac{\partial L_{int}}{\partial \phi},
\end{equation}
where $L_{int}$ contains the infinite vertices involving the interaction of $\phi_1$ with the rest of the fields. 

The following discussion follows closely the methods of \cite{ryder}. The kernel of the wave operator, or equivalently, the Green function $G(y-x)$ of the operator $\square_y$
$$
\square_y G(y-x)=\delta(y-x),
$$
allows to find the following integral relation for $\phi_1$
$$
\int[\phi_1(y)\square_y G(y-x)-G(y-x)\square_y \phi_1(y)] d^4 y =\phi_1(x)-\int  G(y-x)\frac{\partial L_{int}}{\partial \phi(y)} d^4 y.
$$
By assuming that $\phi_1$ vanish fast enough at the spatial infinite, the spatial derivatives in $\square_y$ may be neglected and the last expression becomes
$$
\int[\phi_1(y)\partial_{y_0} G(y-x)-G(y-x)\partial_{y_0} \phi_1(y)] d^4 y =\phi_1(x)-\int  G(y-x)\frac{\partial L_{int}}{\partial \phi(y)} d^4 y.
$$
or equivalently 
$$
\phi_1(x)=\bigg(\int_{y^-_0} -\int_{y^+_0} \bigg)G(y-x)\overleftrightarrow{\partial}_{y_0} \phi_1(y) d^3 y +\int_{y_0^-}^{y_0^+}  G(y-x)\frac{\partial L_{int}}{\partial \phi(y)} d^4 y.
$$
The first two integrals are three dimensional, the first is done at an spatial surface corresponding to the time $y_0^-$at the past of  $x_0$ and the second at another time $y_0^+$ at the future of $x_0$. The last integral is four dimensional, and is performed on the $4$-volume delimited by these two three dimensional surfaces.

At the present point $G(y-x)$ can be any Green function, defined by any possible boundary condition. Particularly  useful examples are the advanced  $\Delta_a(x)$ or retarded $\Delta_r(x)$ Green functions, which vanish for $x_0>0$ and $x_0<0$, respectively, and vanish outside the light cone.  If these functions are employed, then the two spatial integrals in the last formulas are reduced into one.  For instance
$$
\phi_1(x)=\int_{y^-_0}\Delta_r(y-x)\overleftrightarrow{\partial}_{y_0} \phi_1(y) d^3 y +\int  G(y-x)\frac{\partial L_{int}}{\partial \phi(y)} d^4 y.
$$
The first term in the last formula has a simple interpretation. This term is such that
$$
\square_x \int_{y^-_0}\Delta_r(y-x)\overleftrightarrow{\partial}_{y_0} \phi_1(y) d^3 y=\int_{y^-_0}\delta_r(y-x)\overleftrightarrow{\partial}_{y_0} \phi_1(y) d^3 y.
$$
In the limit  $y_0^-\to -\infty$ the last integral vanishes if $x_0$ is fixed, since this point will be always at the future of $y$. Therefore this particular quantity is killed by the action of $\square_x$ and may be identified as a free field, the incoming field. The previous formula, by taking into account the equations of motion \eqref{moushon}, may be expressed after taking this limit as
\begin{equation}\label{inner}
\phi_1(x)=\phi_{1 in}(x) +\int  \Delta_r(x-y)\square_y\phi(y) d^4y.
\end{equation}
Here, the second integral can be done in the full space time volume, as the past time is taken to $-\infty$ and the retarded Green function vanish at the future. Analogously 
\begin{equation}\label{outer}
\phi_1(x)=\phi_{1 out}(x) +\int  \Delta_a(x-y)\square_y\phi(y) d^4y.
\end{equation}
The last expression allows to conclude that
\begin{equation}\label{comuta1}
<0|[\phi_{1in}(x), \phi_{1in}(y)]|0>=<0|[\phi_{1}(x), \phi_{1}(y)]|0>.
\end{equation}
This is seen as follows. The last identity, deduced from \eqref{inner} is true  up to a term proportional to 
$$<0|[\phi_{in}(x),\square_y \phi(y)]|0>.$$ This term vanishes however, due to property $$
<0|\square_y \phi_1(y)|_p>=\square_y e^{ipy} <0|\phi(0)|_p>=-p^2 e^{ipy} <0|\phi(0)|_p>=0.$$
Due to this property the insertion of the identity $\int dp |_p><^p|$ in this extra piece gives vanishing result and the desired identity is proved.

In addition from the mode expansion
 $$
 \phi_1(x) =\int\frac{d^3k}{(2\pi)^{\frac{3}{2}}\omega(k)}[b(k) e^{i\omega(k)t-i k\cdot x}+b^\dag(k) e^{-i\omega(k)t+i k\cdot x}],
 $$
 it is seen that, due to the non standard oscillator algebra of $b(k)$ and $b^\dag(k)$
 $$
[\phi_1(x), \phi_1(y)]=\int \int\frac{d^3k}{(2\pi)^{\frac{3}{2}}\omega(k)}\frac{d^3l}{(2\pi)^{\frac{3}{2}}\omega(l)}\bigg[[b(k), b^\dag(l) ] e^{-i\omega(k)x_0+i k\cdot x} e^{i\omega(l)y_0-i l\cdot y}
$$
$$
-[b(l), b^\dag(k) ] e^{i\omega(k)x_0-i k\cdot x} e^{-i\omega(l)y_0+i l\cdot y}\bigg].
 $$
$$
=- \int\frac{d^3k}{(2\pi)^{\frac{3}{2}}\omega(k)}[e^{-i\omega(k)(x_0-y_0)+i k\cdot (x-y)}-e^{i\omega(k)(x_0-y_0)-i k\cdot (x-y)}]
$$
$$
=i \int\frac{d^3k}{(\pi)^{\frac{3}{2}}\omega(k)}e^{i k\cdot (x-y)}\sin(\omega(k)(x_0-y_0)).
$$
This is consistent with equal time commutator equal to zero, due to the presence of the sinus function in the integral. Define 
\begin{equation}\label{comuta2}
[\phi_1(x), \phi_1(y)]=-i\Delta(x-y),
\end{equation}
where the explicit expression for $\Delta(x-y)$ can be deduced from the last two formulas. Note the unusual minus sign with respect to the textbook standard situation. In QFT books, the function $\Delta(x-y)$ is the same of above, but the commutator has a plus sign. This is the reflection of the non standard commutation relations employed above.  Besides from that, the last function is known to be related to the retarded and advanced Green functions by \begin{equation}\label{retame}\Delta(x-y)=\Delta_a(x-y)-\Delta_r(x-y).\end{equation}
The commutator \eqref{comuta2}  corresponds to free fields. For interacting fields, one has to employ \eqref{comuta1}. However,  the value of $[\phi_{1}(x), \phi_{1}(y)]$ is known to be 
 a c-number \cite{ryder} and therefore \eqref{comuta1} may be replaced by
\begin{equation}\label{comuta3}
[\phi_{1in}(x), \phi_{1in}(y)]=[\phi_{1}(x), \phi_{1}(y)]-i\Delta(x-y),
\end{equation}
in other words, the vacuum expectation value may be deleted.

 By collecting all the above information, consider the fundamental operator for scattering in QFT namely
 $$
Z(J)=<0|I(J)|0>,\qquad I(J)=T e^{i\int \phi(x)J(x)dx},
 $$
 the last quantity satisfies
 $$
-i\frac{\delta I(J)}{\delta J(x)}=T(\phi(x)I(J)).
 $$
 The last formula combined with formulas \eqref{inner} and \eqref{outer} allows to conclude that
 \begin{equation}\label{inner2}
-i\frac{\delta I(J)}{\delta J(x)}=I(x)\phi_{1 in}(x)  -i\int  \Delta_r(x-y)\square_y\frac{\delta I(J)}{\delta J(y)} d^4y,
\end{equation}
\begin{equation}\label{outer2}
-i\frac{\delta I(J)}{\delta J(x)}=\phi_{1 out}(x)I(x) -i\int  \Delta_a(x-y)\square_y\frac{\delta I(J)}{\delta J(y)} d^4y.
\end{equation}
 From the two equations \eqref{inner2} and \eqref{outer2} it is arrived to
 $$
\phi_{1 out}(x) I(x)-\phi_{1 in}(x) I(x)= i\int  \Delta(x-y)\square_y\frac{\delta I(J)}{\delta J(y)} d^4y,
 $$
and this, together with the action of the $S$ matrix defined in \eqref{boca} leads to the following commutator
\begin{equation}\label{hair}
[\phi_{1in}(x), SI(J)]=i\int  \Delta(x-y)\square_y\frac{\delta SI(J)}{\delta J(y)} d^4y.
\end{equation}
In addition, it has been shown in \eqref{comuta3} that the commutator of the fields is a c-number. By applying the following consequence of the Bakker-Campbell formula, which is true only for commutators which are c-numbers, 
$$
[A, e^B]=[A, B] e^B, 
$$
it can be seen that the general solution of the last equation \eqref{hair}
is 
$$
SI(J)=e^{-\int \phi_{1 in}(z)\square_z\frac{\delta}{\delta J(z)}d^4z} F(J),
$$
with $F(J)$ arbitrary.  The proof is found from the version of the Baker Campbell formula given above. To visualize this statement, note that  formula and the last trial function imply
$$
[\phi_{1in}(x), SI(J)]=-\bigg[\int [\phi_{1in}(x), \phi_{1in}(z)] \square_z\frac{\delta}{\delta J(z)}d^4z \bigg]e^{-\int \phi_{1 in}(y)\square_y\frac{\delta}{\delta J(y)}d^4y} F(J)
$$
$$
=ie^{-\int \phi_{1 in}(y)\square_y\frac{\delta}{\delta J(y)}d^4y}\bigg[\int \Delta(x-z) \square_z\frac{\delta F(J)}{\delta J(z)}d^4z \bigg].
$$
This characterizes the left hand side of the equation \eqref{hair}  The right hand side is
$$
i\int  \Delta(x-y)\square_y\frac{\delta SI(J)}{\delta J(y)} d^4y=i\int  \Delta(x-y)\square_ye^{-\int \phi_{1 in}(z)\square_z\frac{\delta}{\delta J(z)}d^4z}\frac {\delta F(J)}{\delta J(y)} d^4y
$$
  The last two formulas are equal, and this shows that \eqref{hair} is satisfied for every $F(J)$. This unknown function is determined by taking into account that, for flat Minkowski space, $S|0>=|0>$, and that
  $$
<0|e^A|0>=<0|:e^A:|0>=1,
  $$
  for any operator $A$. In these terms
$$
<0|IS(J)|0>=<0|F(J)|0>=<0|I(J)|0>=Z(J).
$$ Therefore it is concluded that the scattering operator is given by
\begin{equation}\label{lsz3}
S=:e^{-\int \phi_{1 in}(z)\square_z\frac{\delta}{\delta J(z)}d^4z} :Z(J)\bigg|_{J=0},
\end{equation}
This formula differs by a minus sign in the exponent with respect of the usual situation in QFT. As in the previous section, this analogy has to be not yet taken fully, as the quantity $Z(J)$ may differ from the standard situation. For the full Stelle gravity the last formula is generalized as
$$
S=:e^{\int \Psi_{1\alpha\beta}(z_5)(\square_3+M^2)\frac{\delta}{\delta J_{5\alpha\beta}(z_5)}d^4z_5} ::e^{-\int \Psi_{2 \alpha\beta}(z_4)\square_4\frac{\delta}{\delta J_{4\alpha\beta}(z_4)}d^4z_4} ::e^{\int h_{\alpha\beta}(z_3)\square_3\frac{\delta}{\delta J_{3\alpha\beta}(z_3)}d^4z_3} :
$$
\begin{equation}\label{lsz2}
:e^{\int \phi_{2 in}(z_2)(\square_2+m^2)\frac{\delta}{\delta J_2(z_2)}d^4z_2} ::e^{-\int \phi_{1 in}(z_1)\square_1\frac{\delta}{\delta J_1(z_1)}d^4z_1} :
Z(J_i)\bigg|_{J_i=0}.
\end{equation}
Note that in the last expression several different currents $J_i$ were turned on. Here all the Dirac-Pauli states correspond to negative exponentials and the standard states to positive ones. The last point needed is to characterize this generating functional  $Z(J)$ and the correlation functions it generates.

\section{The generating functional $Z(J)$}
The two formulas derived above for the $S$-matrix are both related to $Z(J)$.  In fact  the reduction formula \eqref{lsz} is related to correlation functions, whose generating functional is $Z(J)$, while the presence of $Z(J)$ in  \eqref{lsz2} is obvious.

The reference \cite{salvio} presents a proposal for $Z(J)$. First of all, as shown in previous sections, Dirac-Pauli variables correspond to states with imaginary eigenvalues $ix$ and $ip$ of the operators $p$
and $x$ whose mean values nevertheless are real. This is a very important point. In this formalism the main formula for going to path integrals is the identity
$$
<^{q_{m+1}}|_{q_{m}}>=\int dp_m <^{q_{m+1}}|_{p_{m}}><^{p_{m}}|e^{iHdt}|_{q_{m}}>,
$$
and, as the recipe employed here requires the standard coordinate and momentum basis, it follows that
$$
<^{q_{m+1}}|_{q_{m}}>=\int dp_m e^{ip_m(q_{m+1}-q_m)-i<H>dt},
$$
where 
$$
<H>=<_p|H|^q>.
$$
This mean value is not the classical hamiltonian, as the Dirac-Pauli variables corresponding to ghost like particles have an imaginary unit $i$ in their definition \cite{salvio}. This prescription has to be applied to those ghost like states. The reference \cite{salvio} then considers the quantization of Stelle gravity in the synchronous gauge in which $g_{00}=1$ and $g_{ij}=0$, where the Dirac Pauli coordinate is $\dot{g}_{ij}$. The value $<^p|H|_q>$ for this model is calculated with the help of the above procedure. Then this theory is generalized to more general gauges and extended to imaginary times $\tau=it$. Perhaps surprisingly, the resulting hamiltonian is nothing less than  the euclidean version of Stelle gravity.  

The above discussion, and in particular the fact that the euclidean Stelle gravity has been suprisingly found, leads to the following recipe for calculating scattering cross sections. 
\\

Step one: find the corresponding correlation function in the euclidean (or imaginary time) path integral formalism, employing the euclidean Stelle action. These result will be renormalizable \cite{dewitt}. Furthermore, if dimensional regularization is employed, the gravitational BRST symmetry may be preserved order by order in $\hbar$ when calculating the corresponding correlations \cite{stelle1}-\cite{stelle2}. 
\\

  Step two: analytically  continue the calculated correlation function  to real time, leading to the desired generating function $Z(J)$.
\\
   
Step three: take the functional in the previous step and apply the formula \eqref{lsz2} for calculating the transition amplitudes and the S matrix. The result will not contain negative norm states and finite cross sections will be found. 
\\

Consider now which functional $Z(J)$ corresponds to the  LSZ fórmula \eqref{lsz} given above. This formula is expressed in terms of the correlation functions $$<0|T\phi_1(y_1)...\phi_1(x_n)|0>.$$ The generating object is 
$$
Z(J)=<0|T e^{i\int J(x)\phi(x)dx}|0>
$$
On the other hand, the discrete version of \eqref{canon} is given by
$$
H=\sum_n\frac{\beta P_1(n) Q_2(n)}{\lambda}-\frac{\lambda^2 P_2(n)^2}{18\gamma}+\frac{\lambda P_2(n)\Delta Q_1(n)}{\beta}-\frac{3}{2}\frac{Q_2(n)^2}{\lambda^2}+\frac{3}{2\beta^2}(Q_1(n+1)-Q_1(n))^2.
$$
As the choice made in the first  sections correspond to $\lambda=1$ and $\beta=i$ it follows that
$$
<p|H|q>=\sum_n iP_1(n) Q_2(n)-\frac{ P_2(n)^2}{18\gamma}-iP_2(n)\Delta Q_1(n)-\frac{3Q_2(n)^2}{2}-\frac{3}{2}(Q_1(n+1)-Q_1(n))^2.
$$
The divergent term $P_1 Q_1$ now is imaginary and do not lead to divergences in the euclidean path integral. This result is consistent with the proposal of \cite{strumia1}-\cite{strumia3}.

The argument above is suggestive since, indirectly, the value $<p|H|q>$ corresponds to the Dirac Pauli prescription of \cite{salvio}, as the ghost like variables are $P_1$ and $Q_1$. The point is that the mean value obtained accidentally corresponds to declare these variables as imaginary. This argument  is valid  for the free part of the hamiltonian, not yet the fully interacting part. On the other hand, the trick of multiplying by $\beta$ imaginary the momenta and by $\beta^{-1}$ the coordinates of the ghost like states also works fine for the quantization in the synchronous gauge in \cite{salvio}. The result, inside the path integral, is equivalent to the Dirac-Pauli prescription these authors present. The recipe for cross section is then the following.
\\

Step one: the same as above.
\\

  Step two: in terms of $Z(J)$ find the euclidean correlation function
   $$<0|T\phi_1(y_1).h_{\alpha\beta}(y_2)..\Psi_{\alpha\beta}(x_3).|0>.$$
  and  analytically  continue the calculated correlation function  to real times.
\\

Step three: take the correlation function calculated in the previous step and apply the formula \eqref{lsz} for calculating the transition amplitudes and the S matrix. Again positive norm states and finite cross sections will be found. 
\\

The only detail that should  be kept  in mind is that one is calculating the transition amplitude corresponding to the usual particles impulses unchanged and ghost like particles with all the signs of the impulses flipped. In other words, if one calculates with the standard textbook LSZ  or with the formula \eqref{lsz} the transition amplitude
$$
A_{if}(p_i, r_i, s_f, t_f),
$$
with $r_i$ and $t_f$ the impulses of the particles created with the non standard creation annihilation operators, then this transition amplitude is really giving 
$$
P_{if}(p_i, -r_i, s_f, -t_f)=A_{if}(p_i, r_i, s_f, t_f)A^\ast_{if}(p_i, r_i, s_f, t_f).
$$
That is, it corresponds to the transition probability with the ghost like modes with flipped impulse signs.
\\
\section{Discussion}
In the present note, LSZ reduction formulas for Stelle gravity were deduced, following a slight simplification of the prescriptions given in \cite{strumia1}-\cite{strumia3}. This required to study the theory in terms of canonical and second quantization, together with the path integral formulation. The resulting $S$ matrix can be renormalized and do not involve negative transition probabilities. Therefore, all the past works about this topic, collected for instance in the books \cite{buchbinder}-\cite{hamber}, which were though as toy calculations for a non consistent model, are revived and can in fact be employed to obtain physically reasonable results. The review \cite{yo} contains a details about renormalization and BRST preservation for these models that may be useful for calculating cross sections.

The advantage of the presentation given here is that the coordinate and momentum operators are obviously hermitian, and avoids a distinction of covariant and contravariant states. This could be useful for a faster visualization. However,  the presentation \cite{strumia1}-\cite{strumia3} is more formal and systematic. The only point is that the hermiticity of certain operators is not obvious at first sight, although at the end, the calculations lead to real eigenvalues and orthogonal eigenvectors, as shown in \eqref{min}.

The classical Stelle model has a well defined Cauchy problem, this was extensively studied in \cite{yo2}-\cite{yo3}. It is a nice result that it also has the properties stated above. If fact, it would be a pity  that a model with such nice classical properties was unstable by quantization. 

The procedures employed here may be applied to large amount of physical theories, not only Stelle gravity but problems of Quantm Mechanics, Solid State Physics or Supergravity and String Theory. Sometimes, the presence of negative norm states is the reason for discarding or making drastic changes in a given model. It is perhaps sometimes convenient to deal with these states and try to figure out a physically reasonable predictions, instead of abandon a model at the very beginning.

An important issue is the measure of the path integral involved. The works \cite{strumia1}-\cite{strumia3} assume that the hamiltonian measure, that is, that the jacobian is the unity in the phase space path integral.
The measure in the lagrangian formalism is then deduced after integrating in the momentum. This measure generates a factor proportional to $\delta^4(0)$ \cite{faddeev} which, in dimensional regularization, is taken equal to zero \cite{stelle1}. In other words, this factor is immaterial for perturbations around \emph{flat} space, as done along the present text. For  perturbations around curved spaces, in particular, for applications such as \cite{gaddam2}-\cite{gaddam3}  this metric may play a fundamental role.  A careful analysis  of the measure and the classification of the symmetries which are broken or not due to its presence is desirable \cite{proeyen}. This issue will be considered in a separate publication.

\section*{Acknowledgements}
The author is supported by CONICET, Argentina and by the Grant PICT 2020-02181.

\end{document}